# Enhanced flux pinning in neutron irradiated MgB$_2$

I.Pallecchi[1], C.Tarantini[1], H.U.Aebersold[2], V.Braccini[1], C.Fanciulli[1], C.Ferdeghini[1], F.Gatti[3], E.Lehmann[2], P.Manfrinetti[4], D.Marré[1], A.Palenzona[4], A.S.Siri[1], M.Vignolo[1] and M. Putti[1]

[1] *INFM-LAMIA/CNR, Dipartimento di Fisica, Via Dodecaneso 33, 16146 Genova, Italy*

[2] *Paul Scherrer Institut, CH-5232 Villagen, Switzerland*

[3] *Dipartimento di Fisica, Via Dodecaneso 33, 16146 Genova, Italy*

[4] *INFM, Dipartimento di Chimica e Chimica Industriale, Via Dodecaneso 31, 16146 Genova, Italy*

**Abstract**

We study the effect of neutron irradiation on the critical current density $J_c$ of isotopically pure polycrystalline Mg$^{11}$B$_2$ samples. For fluences in the range $10^{17}$-$10^{18}$ cm$^{-2}$, $J_c$ is enhanced and its dependence on magnetic field is significantly improved: we demonstrate that, in this regime, point-like pinning centers are effectively introduced in the system proportionally to the neutron fluence. Instead, for larger fluences, a strong suppression of the critical temperature accompanied by a decrease of both the upper critical field $B_{c2}$ and $J_c$ is found.

Since the discovery of superconductivity in $MgB_2$, its 40K transition temperature and moderately low anisotropy have made it interesting for applications in comparison with both low-$T_c$ and high-$T_c$ superconductors. Wide interest has been focused on the study of the physics underlying $MgB_2$ superconducting mechanisms. The upper critical field $B_{c2}$, especially in thin films,[1] can be hugely increased, exceeding that of Nb-based superconductors at all temperatures. On the other hand, the critical current density $J_c$ and its behavior in magnetic field have not yet reached their full potential. The critical current has been thoroughly analyzed in order to establish how its value and magnetic field behavior are correlated with $B_{c2}$ and the flux pinning mechanisms. The important role of grain boundaries as pinning centers has been emphasized.[2] The intentional addition of defects such as nano-particles[3,4] and irradiation damages[5] has been proved to be effective in enhancing both $B_{c2}$ and flux pinning, but a unambiguous discrimination between these two effects has not been formulated yet.

In this paper, we report a quantitative study of the effect of neutron irradiation with increasing fluence from $10^{17}$ to $10^{20}$ cm$^{-2}$ on the critical current density of polycrystalline $MgB_2$ samples. We demonstrate that the grain boundary flux pinning cannot account alone for the enhanced $J_c$ values in magnetic field. On the contrary, we unambiguously show that an additional pinning contribution by point defects is introduced by irradiation.

Seven bars (~1×2×12 mm$^3$) were cut from a $MgB_2$ sample prepared by direct synthesis from Mg and crystalline isotopically enriched $^{11}$B. Each sample was irradiated for a different time at the spallation neutron source SINQ (thermal neutron flux density up to 1.6·10$^{13}$ cm$^{-2}$ s$^{-1}$) at the Paul Sherrer Institut (PSI). The samples were characterized by X-Ray diffraction and the critical temperature $T_c$ was determined by resistivity measurements. The unirradiated clean-limit sample has a resistivity as low as 1.6 μΩcm just above $T_c$, making its defect free structure particularly suitable to study the effectiveness of purposely introduced defects as pinning centers. The magnetization was measured in a Quantum Design SQUID magnetometer up to 5 T. One sample was measured up to 9 T in a PPMS system. Magnetoresistivity was measured in a Quantum Design

PPMS up to 9 T, as well as up to 20.3 T in a resistive magnet at the GHMFL in Grenoble. The full characterization of such samples is reported elsewhere.[6] We want to stress here that the samples whose critical temperature decreases with the neutron fluence down to 9.3 K, present very sharp transitions (see table I), indicating a homogeneous distribution of defects within the samples. Furthermore, we have shown that thermal neutrons rather than fast neutrons are mostly effective in creating damage in the samples, by interaction with the low percentage (lower than 0.5%) of $^{10}$B.

The resistivity of the samples increases almost two orders of magnitude (from 1.6 to 130 µohm·cm) with increasing fluence, indicating the formation of a large number of atomic scale defects. On the other hand, the presence of defects whose size (nearly 5 nm of diameter) matches the $MgB_2$ coherence length has been observed in similarly irradiated samples in reference 7. The upper critical field $B_{c2}$, operatively defined at 90% of the resistive transition, are presented in figure 1 and their values at 5K are reported in table I. It is clear that $B_{c2}$ does not vary appreciably for a fluence of $10^{17}$ cm$^{-2}$ (from 15.4 T to 16.5 T at 5 K), while upon further increasing fluence up to $7.6 \times 10^{17}$ cm$^{-2}$ it strongly increases up to an extrapolated value of 26T at 5K.[8] In a higher fluence regime, the critical temperature is strongly suppressed and $B_{c2}$ is correspondingly lower.

The critical current density is extracted from magnetization hysteresis loops, using the appropriate critical state model.[9] In figure 2, $J_c$ curves as a function of the applied magnetic field at 5K are presented. The critical current density of the unirradiated sample is $2 \cdot 10^9$ A/m$^2$ at 1 Tesla and rapidly decays in magnetic field, becoming negligibly small at 5 Tesla. After neutron irradiation with a fluence in the range $10^{17}$-$10^{18}$ cm$^{-2}$ the critical current density is slightly enhanced, becoming nearly $2.5 \cdot 10^9$ A/m$^2$ at 1 Tesla and, above all, decreasing much more slowly with magnetic field. For the sample P3, the current is still as high as $4 \cdot 10^7$ A/m$^2$ at 8.5 Tesla. This indicates that irradiation, despite suppressing superconductivity as evidenced by the decrease in $T_c$, is an effective means of improving the critical current behavior of $MgB_2$ in magnetic field. With further irradiation, both the critical current and the critical temperature drop, due to the strong suppression of superconductivity by the induced damage.

Our results in the fluence range of $10^{18}$ cm$^{-2}$ appear qualitatively similar to those obtained in neutron irradiated MgB$_2$ bulk samples and wires.[10,11] In ref. 11 the rise of $J_c$ with irradiation was ascribed to the increase of $B_{c2}$, without invoking additional pinning centers. Our evidence is indeed different, maybe due to the higher purity of our pristine sample: at low irradiation levels ($10^{17}$ cm$^{-2}$), despite the upper critical field does not change significantly, the field dependence of $J_c$ is strongly affected by irradiation. In particular, at 5 K and 5 T the critical current densities of the samples P0 and P1 differ by nearly one order of magnitude, while the B$_{c2}$ values differ only by ~7%. In the following we analyze quantitatively all the curves and demonstrate that an additional pinning mechanism must be invoked, directly and unambiguously related to the irradiation fluence.

We employ a percolative model, in order to extract information on the pinning mechanism of flux lines as a function of the increasing irradiation. In this model, described in detail in reference, the sample is considered as made of grains whose orientation is uniformly distributed as sin($\vartheta$) - $\vartheta$ being the angle between the applied magnetic field $B$ and the $c$ axis - and whose upper critical field depends on $\vartheta$ according to the anisotropic Ginzburg-Landau relation: $B_{c2}(\vartheta) = B_{c2}(\pi/2)[\gamma^2 \cos^2(\vartheta) + \sin^2(\vartheta)]^{-1/2}$. Here $\gamma$ is the anisotropy factor of the upper critical field $\gamma = B_{c2}(\pi/2)/B_{c2}(0) = B_{c2}^{//}/B_{c2}^{\perp}$. At a fixed applied field $B$, there is a volume fraction of grains $p$ which are in the normal state, as their orientation $\vartheta$ is such that $B$ exceeds their critical field $B_{c2}(\vartheta)$. As long as the volume fraction $p$ is larger than a critical value $p_c$ a percolative path of superconducting grains exist throughout the sample. The percolation threshold $p_c$ depends on the microstructure and in particular it is determined by the number of neighboring grains of each grain. In our case, the grains have not a regular shape nor they are arranged according to a regular packing, thereby it is not easy to evaluate the exact coordination number $Z$. Considering that for a face centered cubic lattice ($Z=12$) $p_c \sim 0.2$ and for a simple cubic lattice ($Z=6$) $p_c \sim 0.31$ [12], it seems likely that our $p_c$ values are somewhere in between. The critical current density $j_c$ of each grain is determined by their orientation $\vartheta$ and the behavior in magnetic field depends on the particular pinning mechanism.[13] In particular, in granular undoped $MgB_2$, the dominant pinning mechanism is grain boundary pinning, while in irradiated samples also point defects pinning will be considered here. For grain boundaries pinning:

$$j_c = A_{GB} \frac{\left(1 - B/B_{c2}(\vartheta)\right)^2}{\sqrt{B_{c2}(\vartheta) \cdot B}} \qquad A_{GB} = \frac{\mu_0 \cdot S \cdot B_c^2}{2} \qquad (1)$$

where $B_c$ is the thermodynamical critical field, $\mu_0$ the vacuum magnetic permittivity and $S$ the grain boundary surface area per unit volume projected in the direction of the Lorentz force. On the other hand for pinning by point defects:

$$j_c = A_P \frac{\left(1 - B/B_{c2}(\vartheta)\right)^2}{B_{c2}(\vartheta)} \qquad A_P = \frac{\mu_0 \cdot V_t \cdot B_c^2}{2.32 \cdot a} \qquad (2)$$

where $a$ is the average diameter of point defects and $V_t$ is the fraction of flux lines length which lies inside the pinning center. If $L$ is the average distance between pinning centers, for rigid flux lines $V_t=(a/L)^3$, while for flexible flux lines $V_t=(a/L)$. Thermal activation of flux lines is totally neglected in this model which is therefore applicable only at low temperatures.

The fraction $p$ of grains whose orientation is such that their critical current density $j_c$ (equations (1) and/or (2)) is larger than the local current density remain in the superconducting state. The local current density depends on $p$, because the normal grains cannot carry local current densities larger than their own critical current density; thereby the more grains become normal, the more is the additional current that must be carried by the remaining superconducting grains. If the local current density is lower than the smallest grain critical current density (i.e. the critical current density of the grains with $\vartheta \sim 0$) the current flows homogeneously through the whole sample and $p=1$. For increasing local current density, $p$ becomes smaller than $1$, until it eventually reaches $p_c$, at which the macroscopic current density vanishes. The macroscopic current density $J_c$ can be obtained by summing over increasing steps of the applied current density, calculating the local current density and the fraction $p$ at each step. The computation is carried out for each value of external magnetic field $B$, so that the output curve $J_c(B)$ is obtained and compared with the experiment.

The model depends on the following parameters: the anisotropy $\gamma$, the percolation threshold $p_c$, the critical field $B_{c2}^{//}$, the coefficient $A_{GB}$ for grain boundary pinning and the coefficient $A_P$ for point defect pinning. The critical fields $B_{c2}^{//}$ at 5K are experimentally measured, as reported in table I. The values of $p_c$ are fixed to 0.3 for all the samples for simplicity, because for $p_c$ varying between *0.2* and *0.3*, our fitting curves are almost unchanged in the range of magnetic field values where we have experimental data points.

First of all, the unirradiated sample P0 curve is fitted assuming that only the grain boundary contribution to pinning is present. Consistently, the grain boundary nature of the pinning mechanism in the unirradiated sample is indicated by the linearity of the Kramer plot ($J_c^{1/2} \cdot B^{1/4}$ vs. $B/B_{c2}$). The experimental $J_c$ curve (see figure 2) is reproduced for values of the two free parameters $\gamma$=4.4 and $A_{GB}^0 = 7.45 \cdot 10^9$ A/m². At a first sight, considering an additional contribution to pinning for the irradiated samples would increase the number of free parameters for the fit, making the result less reliable. However, we assume that the grain boundary pinning remains unaffected by irradiation, so that the coefficients $A_{GB}$ should obey a scaling law from sample to sample. $A_{GB}$ is proportional to the condensation energy $E_c = \mu_0 B_c^2(T) \approx \mu_0 B_c^2(0)\left(1-(T/T_c)^2\right)^2$ and we assume that $B_c(0) \propto T_c$. Thereby we have the following scaling law:

$$A_{GB} = A_{GB}^0 \frac{T_c^2\left(1-\left(\frac{T}{T_c}\right)^2\right)^2}{T_{c0}^2\left(1-\left(\frac{T}{T_{c0}}\right)^2\right)^2} \tag{3}$$

where $T_{c0}$=39.2 K is the critical temperature of the unirradiated sample and $T$= 5 K.

For the samples irradiated at low fluence, the critical current rescaled by equation (3) lies below the experimental data, indicating that an additional contribution should be considered in order to take into account the measured critical current. For example, in the main panel of figure 3, the experimental critical current of the sample P2 is compared to the rescaled grain boundary contribution, which lies well below. For samples P1, P2 and P3, the rescaled grain boundary contribution to the critical current density is smaller than the experimental data points by a factor 3-4 at 4 Tesla; such discrepancy cannot be accounted for in terms of deviations of the scaling law due, for example, to strain effects[14]. Indeed, defects of nearly 5 nm diameter have been observed in neutron irradiated samples by electron transmission microscopy ; moreover, a downward curvature in the Kramer plots of irradiated samples is observed at low magnetic fields, which is indicative of point defect pinning. The total current is then calculated as the sum of the grain boundary contribution and a point defect contribution, with two free parameters $A_P$ and $\gamma$. In the main panel of figure 3 the point defect contribution and the sum of the two contributions are also plotted, showing a satisfactory agreement with the experimental data. For the other samples the fitting curves with the two contributions are represented as continuous lines in figure 2 and the best fit parameters $A_P$ and $\gamma$ are reported in table 1. The quality of the fit is acceptable for fluences up to $10^{18}$ cm$^{-2}$: reasonably, $A_P$ increases with the fluence, while the anisotropy $\gamma$ decreases, in agreement with literature values[15,11]. Instead, for the samples P4, P5 and P6 the rescaled grain boundary contribution is even larger than the experimental data points, indicating a failure of the scaling procedure, which will be discussed later.

The analysis of the fitting parameters $A_P$ gives us an unambiguous check of the reliability of our description. From equation (2) it can be seen that the parameters $A_P$ should give information on the average distance $L$ between point defect pinning centers induced by irradiation. In the regime of flexible flux lines (pinning point distance larger than the coherence length $\xi_0$), the following proportionality should hold:

$$A_P \propto \frac{1}{L}\mu_0 B_c^2(T) \propto \frac{1}{L}T_c^2\left(1-\left(\frac{T}{T_c}\right)^2\right)^2 \qquad (4)$$

where the scaling of the condensation energy is kept into account. In order to analyze the relationship between the average distance $L$ of point pinning centers and the average distance of defects produced by irradiation, in the inset of figure 3 we plot ($1/L$) estimated from eq. (4) for the samples P1, P2, P3, as a function of (*fluence*)$^{1/3}$. As it can be seen it results in a linear behavior; it is worth noticing that it extrapolates to the origin which represents the unirradiated sample P0. This clearly indicates that the new pinning centers are introduced by irradiation.

The failure of scaling for the highly irradiated samples can be attributed to the rough scaling law that we assume for the condensation energy. In a single band superconductor $E_c(0) = \mu_0 B_c^2(0) \propto N^* \Delta^2(0) \propto N^* T_c^2$, where $N^*$ is the density of state renormalized by the electron-phonon coupling and $\Delta(0)$ is the energy gap. In equation (3) we take into account only the scaling with $T_c$ without considering that in the highly irradiated samples the critical temperature is lowered down to 9 K and this is certainly accompanied by a suppression of $N^*$ (an experimental suppression of $N^*$ in irradiated samples has been indeed observed in reference [16]). Moreover, we have to consider that the less irradiated samples present two gaps while the highly irradiated have probably a single gap, complicating further the evaluation of the condensation energy[17].

Finally we cannot exclude that at the highest fluences a significant volume fraction of the sample might be corrupted by irradiation and the percolation model should be corrected to take it into account.

In summary, we present a systematic analysis of the critical current density and upper critical field of neutron irradiated $MgB_2$ samples, for fluences from $10^{17}$ to $10^{20}$ cm$^{-2}$. There exist two regimes of fluences: in the range $10^{17}$-$10^{18}$ cm$^{-2}$ we measure a significant enhancement of $B_{c2}$ and an improved behavior of $J_c$ in magnetic field; at larger fluences an abrupt suppression of $T_c$, $B_{c2}$ and $J_c$ is found. Thanks to the high purity of our pristine sample, we are able to detect the variations in strength of the pinning mechanisms. We quantitatively demonstrate that in the former regime, the enhancement of $B_{c2}$ alone cannot account for the improved $J_c$. Instead, effective point-like pinning centers are introduced by irradiation proportionally to the fluence, in such a way that the increased pinning force dominates over the suppression of superconductivity associated with damages, resulting in an overall improvement of the critical current density.

**Acknowledgements**

The authors would like to thank Dr. M. Eisterer for his precious help.

**Figure captions:**

**Figure 1:** Upper critical fields as a function of the temperature $T$. The fluences of irradiation in units cm$^{-2}$ indicated in parentheses in the legend are relative to thermal neutrons.

**Figure 2:** Critical current densities extracted at 5K by magnetization measurements; the lines are the calculated curves described in the text, using the parameters listed in table I. In the legend, the fluences of irradiation by thermal neutrons are indicated in parentheses.

**Figure 3:** Main panel: critical current density of the sample P2 and separate contributions to the fitting curves by grain boundaries pinning and point defects pinning. Inset: inverse average distance between point pinning centers in arbitrary units obtained by the fits of the critical current density plotted as a function of the 1/3 power of the fluence. The linear proportionality and the intercept in the origin described by equation (4) are evidenced.

**Table I:** Experimental parameters of the whole set of samples: fluence of neutron irradiation, critical temperature $T_c$, resistive transition width $\Delta T_c$, upper critical field at 5K; in the remaining columns are the results of the fits performed on the critical current density curves: pinning mechanism (grain boundary pinning and/or point defects pinning), multiplicative coefficient for the grain boundary pinning contribution $A_{GB}$ defined in equation (1), multiplicative coefficient for the point defects pinning contribution $A_P$ defined in equation (2), anisotropy factor $\gamma$.

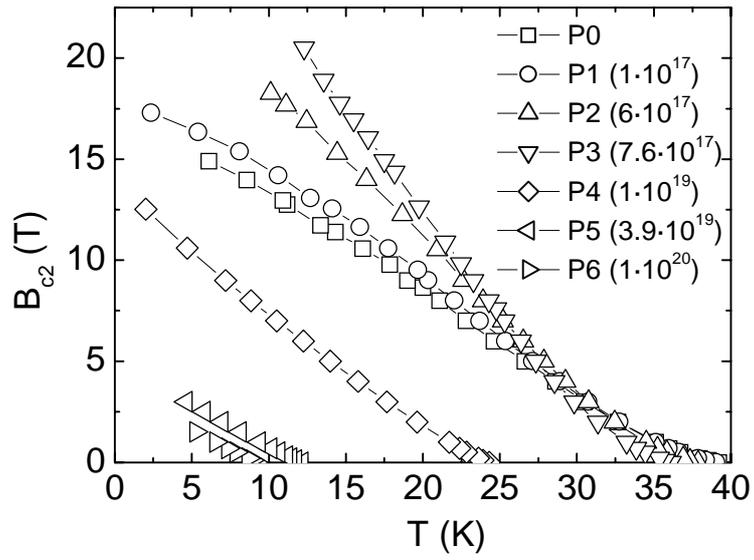

**Figure 1**

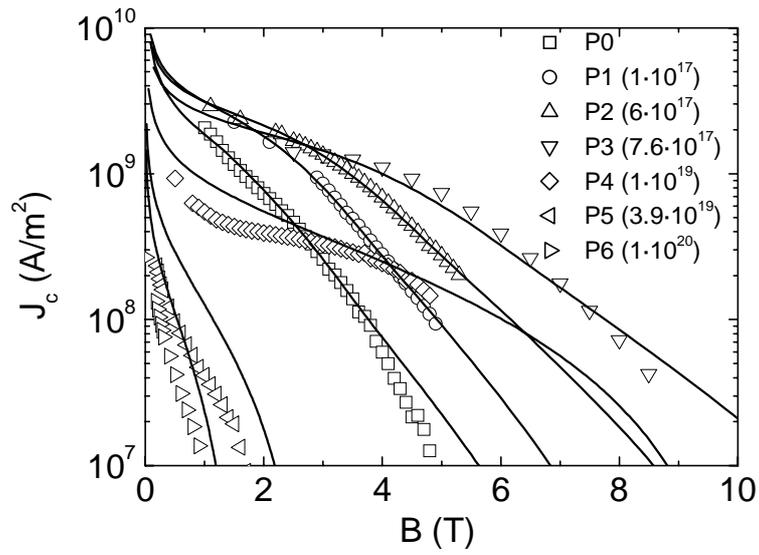

**Figure 2**

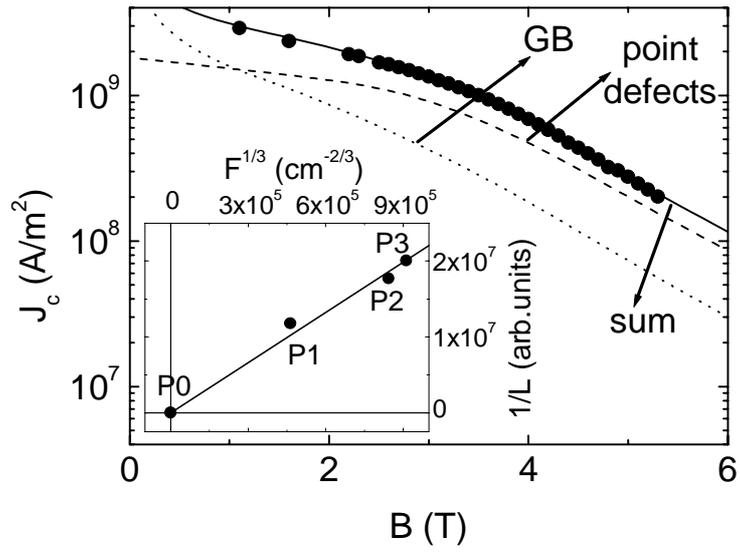

**Figure 3**

**Table I**

| Sample | Fluence (cm$^{-2}$) | $T_c$ (K) | $\Delta T_c$ (K) | $B_{c2}$ at 5K (T) | Pinning mechanism | $A_{GB}$ (A/m$^2$) | $A_P$ (A/m$^2$) | $\gamma$ |
|---|---|---|---|---|---|---|---|---|
| P0 | 0 | 39.2 | 0.2 | 15.4 | GB | 7.45·10$^9$ | | 4.4 |
| P1 | 1·10$^{17}$ | 39.1 | 0.2 | 16.5 | GB + point | 7.40·10$^9$ | 17.4·10$^9$ | 4.4 |
| P2 | 6·10$^{17}$ | 37.9 | 0.2 | * 20.7 | GB + point | 6.94·10$^9$ | 24.5·10$^9$ | 4.4 |
| P3 | 7.6·10$^{17}$ | 36.1 | 0.3 | * 26.2 | GB + point | 6.28·10$^9$ | 25.2·10$^9$ | 4.2 |
| P4 | 1·10$^{19}$ | 24.3 | 0.9 | 10.2 | GB + point | 2.72·10$^9$ | 2.6·10$^9$ | 1 |
| P5 | 3.9·10$^{19}$ | 12.2 | 0.7 | 2.8 | GB | 5.12·10$^8$ | | 1 |
| P6 | 1·10$^{20}$ | 9.2 | 0.2 | 1.6 | GB | 2.11·10$^7$ | | 1 |

The symbol * indicates that the value is not measured, but extrapolated from experimental data.